\documentclass[12pt]{iopart}
\usepackage[dvips]{graphicx}
\usepackage{latexsym}
\usepackage{iopams}
\usepackage{amssymb}
\usepackage{verbatim,times,bbm}

\newcommand{\supa}{\hspace{0.6mm}\epsilon^\sista}
\newcommand{\supb}{\hspace{0.6mm}\epsilon^\sistb}
\newcommand{\osupa}{\hspace{0.6mm}\breve{\epsilon}^\sista}
\newcommand{\osupb}{\hspace{0.6mm}\breve{\epsilon}^\sistb}
\newcommand{\enne}{\mbox{\scriptsize $\mathsf{n}$}}
\newcommand{\ennebis}{\mbox{\footnotesize $\mathsf{n}$}}
\newcommand{\indn}{i_{\enne}}
\newcommand{\un}{\mbox{\tiny $1$}}
\newcommand{\undu}{\mbox{\tiny $1,\hspace{-0.4mm}2$}}
\newcommand{\sosp}{\mbox{\tiny $\hspace{-0.2mm},\hspace{-0.4mm}...,\hspace{0.1mm}$}}
\newcommand{\sista}{{\hspace{0.3mm}\mbox{\tiny $\mathsf{A}$}}}
\newcommand{\sistb}{{\hspace{0.3mm}\mbox{\tiny $\mathsf{B}$}}}
\newcommand{\sistaa}{{\mbox{\tiny $\mathsf{A}$}}}
\newcommand{\sistbb}{{\mbox{\tiny $\mathsf{B}$}}}
\newcommand{\hilba}{{\mathcal{H}_{\mbox{\tiny $\mathsf{A}$}}}}
\newcommand{\hilbb}{{\mathcal{H}_{\mbox{\tiny $\mathsf{B}$}}}}
\newcommand{\na}{{N_{\mbox{\tiny $\mathsf{A}$}}}}
\newcommand{\nb}{{N_{\mbox{\tiny $\mathsf{B}$}}}}

\newcommand{\rea}{{\mbox{\tiny $\mathrm{R}$}}}
\newcommand{\phan}{{\phantom{\ast}}}
\newcommand{\pha}{{\phantom{\rea}}}
\newcommand{\phant}{{\phantom{\dagger}}}
\newcommand{\dd}{{\mathsf{d}}}

\newcommand{\rr}{{\mathsf{r}}}
\newcommand{\RR}{{\mathsf{R}}}
\newcommand{\hhilb}{{\hat{\mathcal{H}}}}
\newcommand{\hhilba}{{\hat{\mathcal{H}}_{\mbox{\tiny $\mathsf{A}$}}}}
\newcommand{\hhilbb}{{\hat{\mathcal{H}}_{\mbox{\tiny $\mathsf{B}$}}}}
\newcommand{\bas}{{\hat{\mathrm{e}}}}
\newcommand{\hrho}{{\hat{\rho}}}
\newcommand{\rema}{{\hspace{0.3mm}\mathtt{RM}\hspace{0.3mm}}}
\newcommand{\dens}{\mathcal{D}(\mathcal{H})}
\newcommand{\densa}{\mathcal{D}(\hilba)}
\newcommand{\densb}{\mathcal{D}(\hilbb)}

\newcommand{\ide}{\hat{\mathbb{I}}}
\newcommand{\ida}{\hat{\mathbb{I}}^{\mbox{\tiny $\mathsf{A}$}}}
\newcommand{\idb}{\hat{\mathbb{I}}^{\mbox{\tiny $\mathsf{B}$}}}

\newcommand{\eau}{\mathfrak{E}_1^\sista}
\newcommand{\eadu}{\mathfrak{E}_2^\sista}
\newcommand{\ebu}{\mathfrak{E}_1^\sistb}
\newcommand{\ebdu}{\mathfrak{E}_2^\sistb}
\newcommand{\ean}{\mathfrak{E}_{\ennebis}^\sista}
\newcommand{\ebn}{\mathfrak{E}_{\ennebis}^\sistb}
\newcommand{\eak}{\mathfrak{E}_k^\sista}
\newcommand{\eal}{\mathfrak{E}_l^\sista}
\newcommand{\ebk}{\mathfrak{E}_k^\sistb}
\newcommand{\ebl}{\mathfrak{E}_{l}^\sistb}
\newcommand{\mara}{\hrho_\sista}
\newcommand{\marb}{\hrho_\sistb}
\newcommand{\tra}{\tr_\sistaa}
\newcommand{\trb}{\tr_\sistbb}
\newcommand{\ope}{\hrho(\mathfrak{E}_{\undu}^{\sista,\sistbb})}
\newcommand{\op}{\hrho(\mathfrak{E}_{\un\sosp\enne}^{\sista,\sistbb})}
\newcommand{\abi}{(\mbox{\footnotesize $\mathsf{A}\rightarrow\mathsf{B}$})}
\newcommand{\coco}{\mathrm{c.c.}}
\newcommand{\xau}{\hat{X}^\sista_1}
\newcommand{\xadu}{\hat{X}^\sista_2}
\newcommand{\xal}{\hat{X}^\sista_l}
\newcommand{\xan}{\hat{X}^\sista_\mathsf{n}}
\newcommand{\xbu}{\hat{X}^\sistb_1}
\newcommand{\xbdu}{\hat{X}^\sistb_2}
\newcommand{\xbl}{\hat{X}^\sistb_l}
\newcommand{\xbn}{\hat{X}^\sistb_\mathsf{n}}
\newcommand{\yau}{\hat{Y}^\sista_1}
\newcommand{\yadu}{\hat{Y}^\sista_2}
\newcommand{\yal}{\hat{Y}^\sista_l}
\newcommand{\yan}{\hat{Y}^\sista_\mathsf{n}}
\newcommand{\ybu}{\hat{Y}^\sistb_1}
\newcommand{\ybdu}{\hat{Y}^\sistb_2}
\newcommand{\ybl}{\hat{Y}^\sistb_l}
\newcommand{\ybn}{\hat{Y}^\sistb_\mathsf{n}}
\newcommand{\oper}{\hrho(\hat{X}_{1,\ldots,\mathsf{n}}^{\sista,\sistbb},\hat{Y}_{1,\ldots,\mathsf{n}}^{\sista,\sistbb})}
\newcommand{\xak}{\hat{X}^\sista_k}
\newcommand{\yak}{\hat{Y}^\sista_k}
\newcommand{\xbk}{\hat{X}^\sistb_k}
\newcommand{\ybk}{\hat{Y}^\sistb_k}
\newcommand{\aaa}{\hat{A}}
\newcommand{\bbb}{\hat{B}}
\newcommand{\hsi}{\hat{\sigma}}

\newcommand{\opu}{\hrho(\mathfrak{E}_{\un}^{\sista,\sistbb})}

\newcommand{\vhrho}{{\hat{\varrho}}}
\newcommand{\vmara}{\vhrho_\sista}
\newcommand{\vmarb}{\vhrho_\sistb}
\newcommand{\vope}{\vhrho(\mathfrak{E}_{\undu}^{\sista,\sistbb})}

\newcommand{\calida}{{\mathfrak{I}}^{\hspace{0.3mm}\mbox{\tiny $\mathsf{A}$}}}
\newcommand{\calidb}{{\mathfrak{I}}^{\hspace{0.3mm}\mbox{\tiny $\mathsf{B}$}}}

\newcommand{\caltrb}{{\mathfrak{T}}^{\hspace{0.4mm}\mbox{\tiny $\mathsf{B}$}}}
\newcommand{\hrhotb}{{\hat{\rho}}^{{\mathsf{T}}_{\hspace{-0.3mm}\mbox{\tiny $\mathsf{B}$}}}}
\newcommand{\marbt}{{\hrho_\sistb}^{{\mathsf{T}}}}
\newcommand{\phanto}{{\phantom{\mathsf{T}}}}

\newcommand{\filta}{{\hat{F}^\sista}}
\newcommand{\filtb}{{\hat{F}^\sistb}}

\newtheorem{proposition}{Proposition}

\newtheorem{theorem}{Theorem}
\newtheorem{corollary}{Corollary}

\begin{document}

\title{A class of inequalities inducing new separability criteria for bipartite quantum systems}

\author{Paolo Aniello$^{\hspace{0.3mm}\ast\,\ddagger}$ and Cosmo Lupo$^{\hspace{0.3mm}\ast\,\dagger}$}

\address{$^\ast$ Dipartimento di Scienze Fisiche dell'Universit\`a di Napoli ``Federico II'', and INFN -- Sezione
di Napoli, via Cintia I-80126 Napoli, Italy \\
$^\ddagger$ Facolt\`a di Scienze Biotecnologiche, Universit\`a di
Napoli ``Federico II''\\$^\dagger$ Research Center for Quantum
Information, Slovak Academy of Sciences, D\'ubravsk\'a cesta~9, 845
11 Bratislava, Slovakia }

\eads{\mailto{aniello@na.infn.it}, \mailto{lupo@na.infn.it}}

\begin{abstract}
Inspired by the realignment or computable cross norm criterion, we
present a new result about the characterization of quantum
entanglement. Precisely, an interesting class of inequalities
satisfied by all separable states of a bipartite quantum system is
derived. These inequalities induce new separability criteria that
generalize the realignment criterion.
\end{abstract}


\section{Introduction}
\label{intro}

Entanglement is a very peculiar and essential feature of quantum
theory, as recognized since the early stages of development of the
theory by Einstein, Podolsky and Rosen~\cite{Einstein}, and by
Schr\"odinger~\cite{Schr1,Schr2} (who introduced the german term
`Verschr\"ankung' and translated it into English as `entanglement').
Recently, entanglement has been investigated with a renewed interest
motivated by ideas and applications stemming from the field of
quantum information science~\cite{NiCh}. Indeed, nowadays quantum
entanglement is not only regarded as a fundamental key for the
interpretation of quantum mechanics but also as an important
resource for quantum information, communication and computation
tasks~\cite{Bouwmeester}. However, despite the great efforts made by
the scientific community in the past decades, there are still
several open issues regarding the mathematical characterization of
entangled quantum states: the study of \emph{bipartite} entanglement
is already a difficult, extensive and rapidly evolving subject, even
in the case of quantum systems with a finite number of levels, and
\emph{multipartite} entanglement is still a rather obscure matter
(see, for instance, the review papers~\cite{Bruss,Horos} and
references therein). In the present contribution, our discussion
will be restricted to the case of bipartite systems with a finite
number of levels.

According to the definition due to Werner~\cite{Werner},
\emph{entangled} (mixed) states differ from \emph{separable} states
since they cannot be prepared using only local operations and
classical communication; hence, they may exhibit non-classical
correlations. In mathematical terms, a mixed state $\hrho$ --- a
positive (trace class) operator of unit trace --- in a composite
Hilbert space $\mathcal{H}=\hilba\otimes\hilbb$ is called
\emph{separable} if it can be represented as a convex sum of product
states (the sum converging with respect to the trace norm):
\begin{equation} \label{separable}
\hrho = \sum_i p_i\; \hrho_i^\sista \otimes \hrho_i^\sistb,
\end{equation}
with $p_i > 0$ and $\sum_i p_i = 1$; otherwise, $\hrho$ is said to
be \emph{nonseparable} or \emph{entangled}. We remark that, if
$\hrho$ is a separable state, decomposition~{(\ref{separable})}
--- a \emph{separability decomposition} of the state $\hrho$
--- is in general not unique, and it can be assumed to be a
\emph{finite} sum if the Hilbert space $\mathcal{H}$ is
finite-dimensional. The smallest number of terms in the sum (usually
called \emph{cardinality} of the separable state $\hrho$) is not
larger than the squared dimension of the \emph{total} Hilbert space
of the system $\mathcal{H}$ (see~\cite{Horo}).

Since quantum entanglement, as already mentioned, is a very
important subject (also in view of its potential applications),
separability criteria are extremely precious tools. As far as we
know, separability criteria found so far fall into two classes: on
one hand, criteria based on both necessary and sufficient
conditions, but not practically implementable; on the other hand,
criteria that are relatively easy to apply, but rely on only
necessary, or only sufficient, conditions. Among necessary and
sufficient criteria, we mention the `positive maps
criterion'~\cite{Horo96}, which leads to the `Peres-Horodecki
criterion'~\cite{Horo96,Peres96} for $2\times 2$ and $2\times 3$
bipartite systems (in these special cases, we have an operational
necessary and sufficient criterion), and the `contraction
criterion'~\cite{3H}. Among necessary conditions for separability
(or, equivalently, sufficient conditions for entanglement), it is
worth mentioning the celebrated `PPT criterion'~\cite{Peres96}
(which, in the special cases of $2\times 2$ and $2\times 3$
bipartite systems, is precisely the Peres-Horodecki
necessary-sufficient criterion), the `reduction
criterion'~\cite{Horo99,Cerf}, the `majorization
criterion'~\cite{Nielsen}, and the criterion that was proposed in
ref.~\cite{ReCr} with the name of `realignment criterion' (RC) and
in ref.~\cite{CCN} with the name of `computable cross norm
criterion', which will be central in the present contribution.

In this paper, we reconsider the RC and, using essentially the same
tools that allow to prove this criterion, we obtain a new class of
inequalities satisfied by all separable states of a bipartite
quantum system. These inequalities potentially induce new
separability criteria; i.e.\ a certain inequality produces a
separability criterion if it makes sense to check it for a generic
state and it is actually violated by some state. Every state that
violates such an inequality is then nonseparable. Numerical
calculations show that from the class of inequalities introduced in
the present paper one obtains, indeed, a wide class of new
separability criteria. This class contains, in particular, the
original RC and a powerful separability criterion which is the main
result obtained in ref.~{\cite{Zhang}}. Other remarkable particular
cases are given by the `enhancement by local filtering operations'
of the RC and of the criterion of ref.~{\cite{Zhang}}, see
Sect.~{\ref{conclusions}}. All the new criteria share with the RC
the important property of being easily implementable, and they are,
in general, independent of the original RC. For instance, it can be
shown that the criterion derived in ref.~{\cite{Zhang}} is stronger
than the standard RC.

The paper is organized as follows. In Sect.~\ref{fromSchmidt}, the
RC is reviewed starting from the Schmidt decomposition of pure
states. In Sect.~\ref{new}, by an argument similar to the one that
allows to prove the RC, we derive the announced class of
inequalities. As mentioned before, this class of inequalities induce
new separability criteria; a few simple examples of entanglement
detection are given in Sect.~{\ref{examples}}. Finally, in
Sect.~\ref{conclusions}, conclusions are drawn.

\section{From the Schmidt decomposition to the realignment criterion}
\label{fromSchmidt}

Aim of the present section is to review some known facts about the
\emph{realignment criterion} (RC) for bipartite quantum systems. We
will try, in particular, to highlight the relation between the
\emph{Schmidt decomposition}~\cite{Schmidt} of a bipartite (pure or
mixed) state and the RC. This relation will be our starting point
for establishing novel separability criteria related to the RC.

Let us consider a bipartite quantum system with carrier Hilbert
space $\mathcal{H} = \hilba \otimes \hilbb$, with $\hilba \cong
\mathbbm{C}^{\na}$ and $\hilbb \cong \mathbbm{C}^{\nb}$, where we
assume that $2\le\na,\nb<\infty$. We will fix in the `local Hilbert
spaces' $\hilba,\hilbb$ orthonormal bases $\{ |n\rangle \}_{n=1,
\dots ,\na}$ and $\{ |\nu\rangle \}_{\nu=1, \dots , \nb}$,
respectively (notice that we use Latin indexes for the subsystem
$\mathsf{A}$ and Greek indexes for the subsystem $\mathsf{B}$).
Then, a state vector $|\psi\rangle\in\mathcal{H}$ of the composite
system can be written as $|\psi\rangle = \sum_{n,\nu} \psi_{n\nu}
|n\rangle|\nu\rangle$ (for notational conciseness, we will
occasionally omit the tensor product symbol). It is clear that one
can regard the components of the vector $|\psi\rangle$ in the given
basis as the entries of a $\na \times \nb$ matrix $\psi$:
\begin{equation} \label{pure_R}
\psi \equiv \left[\psi_{n\nu}\right];
\end{equation}
here notice that the Latin and the Greek indexes play respectively
the role of row and column indexes of the matrix $\psi$.

For instance, in the simplest case of a two-qubit system --- i.e.\
$\hilba \cong \hilbb \cong \mathbbm{C}^2$ --- with a state vector
\begin{equation}\label{2_qubits_pure}
|\psi\rangle = \psi_{11}|11\rangle + \psi_{12}|12\rangle +
\psi_{21}|21\rangle + \psi_{22}|22\rangle\in \hilba \otimes \hilbb
\end{equation}
is associated a $2\times 2$ the matrix
\begin{eqnarray}\label{2_qubits_pure_R}
\psi = \left[\begin{array}{cc} \psi_{11} & \psi_{12} \\
\psi_{21} & \psi_{22} \end{array}\right].
\end{eqnarray}

As it is well known, a Schmidt decomposition (non-uniquely
determined) of the state vector $|\psi\rangle$ comes from a singular
value decomposition (SVD)~\cite{Horn} of the matrix of
coefficients~(\ref{pure_R}):
\begin{equation} \label{svdec}
\psi = U\, \Delta\, V= \left[\psi_{n\nu}=\sum_{m,\mu} U_{nm}\,
\Delta_{m\mu} V_{\mu\nu}\right]
\end{equation}
where $U$ and $V$ are respectively $\na \times \na$ and $\nb \times
\nb$ unitary matrices, while $\Delta$ is a $\na\times\nb$ matrix
with non-negative real entries and, precisely, with the only
non-vanishing entries along the principal diagonal. Setting
$\dd\equiv \min \{\na, \nb \}$,
one can choose the unitary matrices $U,V$ in such a way that
the diagonal entries $(\delta_1, \delta_2, \dots \delta_\dd)$ of
$\Delta$
--- the `singular values' of $\psi$ --- are arranged in non-increasing
order: $\delta_1 \geq \delta_2 \ge \cdots \ge\delta_\dd$ (we will
always follow this convention for the singular values). The
SVD~{(\ref{svdec})} allows to write the state vector $|\psi\rangle$
in the Schmidt canonical form:
\begin{equation}
|\psi\rangle = \sum_{k=1}^\rr \delta_k\, |\phi^\sista_k\rangle
|\phi^\sistb_k\rangle,\ \ \ \rr:= \max\left\{k\in\{1,\ldots,\dd\}:\
\delta_k >0\right\},
\end{equation}
where $\rr$ is the \emph{Schmidt rank} of the vector $|\psi\rangle$
(observe that $\rr=\mathrm{rank}(\psi)$), and the sets of vectors
$\{|\phi^\sista_k\rangle \}_{k=1}^{\rr}$ and $\{
|\phi^\sistb_k\rangle \}_{k=1}^{\rr}$ are orthonormal systems,
respectively in $\hilba$ and $\hilbb$, determined by the unitary
matrices $U$ and $V$; indeed, we have that
\begin{equation}
|\psi\rangle  = \sum_{n,\nu} \psi_{n\nu} |n\rangle|\nu\rangle =
\sum_{n,\nu} \sum_{m,\mu} U_{nm}\, \Delta_{m\mu} V_{\mu\nu}\,
|n\rangle|\nu\rangle ,
\end{equation}
hence:
\begin{equation}
|\phi^\sista_k\rangle= \sum_{n=1}^{\na} U_{nk}\,  |n\rangle, \ \ \
|\phi^\sistb_k\rangle= \sum_{\nu=1}^{\nb} V_{k\nu}\, |\nu\rangle ,\
\ \ k=1,\ldots , \rr.
\end{equation}
Notice that the positive real Schmidt coefficients $\{ \delta_k
\}_{k=1}^\rr$ are only constrained by the normalization of the state
vector $|\psi\rangle$ to fulfill the condition: $\sum_k \delta_k^2 =
1$. We stress also that, although the matrix $\psi$ depends on the
choice of the orthonormal bases $\{ |n\rangle \}_{n=1, \dots ,\na}$
and $\{ |\nu\rangle \}_{\nu=1, \dots , \nb}$, its singular values
--- i.e. the Schmidt coefficients of $|\psi\rangle$ --- are
basis-independent.

The separability of pure states is related with the Schmidt
coefficients: separable states have only a single non-vanishing
Schmidt coefficient, i.e.\ $(\delta_1,\delta_2,\ldots,\delta_\dd) =
(1, 0, \ldots, 0)$ (in this case, $\rr=\mathrm{rank}(\psi)=1$); on
the other hand, maximally entangled pure states have Schmidt
coefficients $(\delta_1,\delta_2,\ldots,\delta_\dd)=
\big(1/\sqrt{\dd}, 1/\sqrt{\dd}, \dots
1/\sqrt{\dd}\hspace{0.3mm}\big)$ (in this case, $\rr=\dd$). In the
general case, one can consider the Schmidt rank
$\rr=\mathrm{rank}(\psi)$ as an entanglement estimator (see
refs.~\cite{Schmidt_m,Eisert,Lupo} for extensions of this approach
to density operators).

In order to highlight the link between the Schmidt decomposition and
the RC, it is convenient to describe the pure state $|\psi\rangle$
as a density operator (precisely, as a rank-one projector):
\begin{equation}
\hat{\rho}_\psi^\pha= |\psi\rangle\langle\psi| = \sum_{m,\mu,n,\nu}
\psi_{m\mu}^\phan \psi_{n\nu}^*\, |m\rangle|\mu\rangle\langle
n|\langle\nu |;
\end{equation}
hence, with respect to the fixed orthonormal basis
$\{|n\rangle|\nu\rangle\}$ in $\mathcal{H}$, the pure state
$\hat{\rho}_\psi^\pha$ is now identified by the $(\na \nb) \times
(\na \nb)$ square matrix
\begin{equation}
\rho_\psi^\pha=\left[{\rho_\psi^\pha}_{(m\mu)(n\nu)}\right],\ \ \
{\rho_\psi^\pha}_{(m\mu)(n\nu)} =\langle m|\langle\mu
|\hspace{0.4mm}\hat{\rho}_\psi^\pha|n\rangle|\nu\rangle=\psi_{m\mu}^\phan\psi_{n\nu}^*,
\end{equation}
rather than by the $\na\times\nb$ matrix $\psi$. Here the indexes
$(m\mu)$ and $(n\nu)$ have to be regarded as \emph{double indexes};
explicitly: $(m\mu) \leftrightarrow \nb (m-1) + \mu$ and $(n\nu)
\leftrightarrow \nb (n-1) + \nu$.

With the square matrix $\rho_\psi^\pha$ one can associate a
\emph{realigned} matrix $\rho_\psi^\rea$, which is a $N_\sistaa^2
\times N_\sistbb^2$ rectangular matrix with entries:
\begin{equation}
{\rho_\psi^\rea}_{(mn)(\mu\nu)} := \psi_{m\mu}^\phan \psi_{n\nu}^* =
{\rho_\psi^\pha}_{(m\mu)(n\nu)},
\end{equation}
where the double indexes $(mn) \leftrightarrow \na (m-1) + n$ and
$(\mu\nu) \leftrightarrow \nb (\mu -1) + \nu$ refer respectively to
rows and columns of the matrix $\rho_\psi^\rea$. Thus, with the
density matrix $\rho_\psi^\pha$, is associated the realigned matrix
$\rho_\psi^\rea$ having precisely the same entries, but
\emph{arranged in a different way}. It is immediate to check that
--- denoting by the symbol $\odot$ the \emph{Kronecker product} of matrices
(in order to avoid confusion with the tensor product $\otimes$ of
the subsystems $\mathsf{A}$ and $\mathsf{B}$) and by $M^*$ the
\emph{complex conjugate} of a matrix $M$ (rather than the adjoint,
which will be denoted by $M^\dagger$)
--- the following relation holds
\begin{equation}
\rho_\psi^\rea = \psi \odot \psi^*.
\end{equation}
Hence, considering the SVD~{(\ref{svdec})} of the matrix $\psi$, and
using the mixed product property of the Kronecker product, we obtain
the following SVD of the realigned matrix:
\begin{equation}\label{pure_SVD}
\rho_\psi^\rea = \mathcal{U}\, \Lambda\, \mathcal{V},\ \ \
\mbox{with:}\ \ \mathcal{U} = U \odot U^*,\ \Lambda=\Delta \odot
\Delta,\ \mathcal{V}=V \odot V^*,
\end{equation}
where we have taken into account that $\Delta=\Delta^*$. Therefore,
if the Schmidt coefficients of the state vector $|\psi\rangle$ are
$\{ \delta_k \}_{k=1}^\dd$, then the associated realigned matrix
$\rho_\psi^\rea$ has singular values coinciding with the principal
diagonal entries of the matrix $\Delta \odot \Delta$; namely: $\{
\lambda_{(hk)} = \delta_h \delta_k \}$, where $(hk) \leftrightarrow
\dd(h-1)+k$. Notice that the rank of the realigned matrix
$\rho_\psi^\rea$ is given by: $\RR=\mathrm{rank}(\Delta \odot
\Delta)=\mathrm{rank}(\Delta)^2=\mathrm{rank}(\psi)^2\equiv\rr^2$.
Moreover, denoted by $\|\cdot\|_{\tr}$ the trace norm, we have that
\begin{eqnarray}
\|\rho_\psi^\rea\|_{\tr} := \tr(|\rho_\psi^\rea|) & = &
\tr\!\left(((\rho_\psi^\rea)^\dag
\rho_\psi^\rea)^{\frac{1}{2}}\right) \nonumber\\
& = & \sum_{h,k=1}^{\dd} \lambda_{(hk)}= \sum_{h,k=1}^{\dd} \delta_h
\delta_k = 1 + \sum_{h \neq k} \delta_h \delta_k \, ;
\end{eqnarray}
hence:
\begin{proposition}
Let $\hrho_\psi=|\psi\rangle\langle\psi|$ be a pure state in
$\mathcal{H}=\hilba\otimes\hilbb\,$. Then, the associated realigned
matrix $\rho_\psi^\rea = \psi \odot \psi^*$ satisfies:
\begin{equation} \label{inequa}
\|\rho_\psi^\rea\|_{\tr} = \sum_{h,k=1}^{\dd} \lambda_{(hk)}=1 +
\sum_{h \neq k} \delta_h \delta_k\ge 1 .
\end{equation}
Thus, inequality~{(\ref{inequa})} is saturated if and only if the
pure state $\hrho_\psi$ is separable:
\begin{equation}
\|\rho_\psi^\rea\|_{\tr} = 1\ \Leftrightarrow \
|\psi\rangle=|\psi^\sista\rangle\otimes|\psi^\sistb\rangle.
\end{equation}
\end{proposition}

For example, in the case of a two-qubit system, the density matrix
$\rho_\psi^\pha$ associated with the state
vector~{(\ref{2_qubits_pure})} has the following form:
\begin{eqnarray}
\rho_\psi^\pha = \left[\begin{array}{cccc}
\psi_{11}\psi_{11}^* & \psi_{11}\psi_{12}^* & \psi_{11}\psi_{21}^* & \psi_{11}\psi_{22}^* \\
\psi_{12}\psi_{11}^* & \psi_{12}\psi_{12}^* & \psi_{12}\psi_{21}^* & \psi_{12}\psi_{22}^* \\
\psi_{21}\psi_{11}^* & \psi_{21}\psi_{12}^* & \psi_{21}\psi_{21}^* & \psi_{21}\psi_{22}^* \\
\psi_{22}\psi_{11}^* & \psi_{22}\psi_{12}^* & \psi_{22}\psi_{21}^* &
\psi_{22}\psi_{22}^* \end{array}\right].
\end{eqnarray}
The corresponding realigned matrix is
\begin{eqnarray}\label{2_qubits_pure_R_I}
\rho_\psi^\rea & = & \left[\begin{array}{cccc}
\psi_{11}\psi_{11}^* & \psi_{11}\psi_{12}^* & \psi_{12}\psi_{11}^* & \psi_{12}\psi_{12}^* \\
\psi_{11}\psi_{21}^* & \psi_{11}\psi_{22}^* & \psi_{12}\psi_{21}^* & \psi_{12}\psi_{22}^* \\
\psi_{21}\psi_{11}^* & \psi_{21}\psi_{12}^* & \psi_{22}\psi_{11}^* & \psi_{22}\psi_{12}^* \\
\psi_{21}\psi_{21}^* & \psi_{21}\psi_{22}^* & \psi_{22}\psi_{21}^* &
\psi_{22}\psi_{22}^*
\end{array}\right]
\nonumber \\
& = &\left[\begin{array}{cc} \psi_{11} & \psi_{12} \\ \psi_{21} &
\psi_{22} \end{array}\right] \odot \left[\begin{array}{cc}
\psi_{11}^* & \psi_{12}^* \\ \psi_{21}^* & \psi_{22}^*
\end{array}\right]=\psi\odot\psi^*.
\end{eqnarray}
Hence, if the singular values of the matrix $\psi$ are $(\delta_1,
\delta_2)$, the realigned matrix~{(\ref{2_qubits_pure_R_I})} has
singular values $(\delta_1^2\ge \delta_1 \delta_2= \delta_2
\delta_1\ge \delta_2^2)$. If the Schmidt rank of the state vector
$|\psi\rangle$ is $\rr$, then the rank of the realigned matrix
$\rho_\psi^\rea$ is $\RR=\rr^2\in\{1,4\}$. If $\hat{\rho}_\psi^\pha$
is a \emph{separable} pure state, then $\rho_\psi^\rea$ has only one
non-vanishing singular value, i.e.\ $\{\lambda_{(hk)}\} = (1, 0, 0,
0)$ and $\RR=1$; moreover: $\|\rho_\psi^\rea\|_{\tr}=1$. If,
otherwise, $\hat{\rho}_\psi^\pha$ is entangled, then $\RR=4$ and
$\|\rho_\psi^\rea\|_{\tr}> 1$; in particular, for a maximally
entangled pure state we have: $\{\lambda_{(hk)}\} = (1/2, 1/2,
1/2,1/2)$ and $\|\rho_\psi^\rea\|_{\tr}= 2$.

Having established a relation between the separability of pure
states and the singular values of the associated realigned matrices,
we now proceed to extend this relation to a \emph{generic} density
operator in $\mathcal{H} = \hilba \otimes \hilbb$.

The first step is to observe that the association of a realigned
matrix with a density operator is a straightforward generalization
of the association introduced for pure states. Let $\hat{\rho}$ be a
density operator in $\hilba \otimes \hilbb$ ($\hat{\rho}\ge 0$,
$\tr(\hat{\rho})=1$), and let $\rho$ denote the corresponding
density matrix with respect to the fixed product basis
$\{|n\rangle|\nu\rangle\}$ in $\hilba \otimes \hilbb$:
\begin{equation}
\rho =\left[\rho_{(m\mu)(n\nu)}^\pha=\langle m|\langle\mu
|\hspace{0.4mm}\hat{\rho}\hspace{0.4mm}|n\rangle|\nu\rangle\right],
\end{equation}
where $(m\mu) \leftrightarrow \nb (m-1) + \mu$ and $(n\nu)
\leftrightarrow \nb (n-1) + \nu$. Then, as above, one can associate
a $N_\sista^2\times N_\sistb^2$ realigned matrix $\rho^\rea$ with
the $\na\nb\times\na\nb$ density matrix $\rho$ in the following way:
\begin{equation} \label{relrea}
\rho_{(mn)(\mu\nu)}^\rea := \rho_{(m\mu)(n\nu)}^\pha.
\end{equation}
It is clear that the association $\hat{A}\mapsto A^\rea$ (relative
to the fixed product basis in $\hilba \otimes \hilbb$) is actually
well defined for \emph{any} linear operator $\hat{A}$ in
$\mathcal{H}=\hilba \otimes \hilbb$.

For example, in the case of a two-qubit system, with a density
matrix
\begin{eqnarray}
\rho = \left[\begin{array}{cccc}
\rho_{11} & \rho_{12} & \rho_{13} & \rho_{14} \\
\rho_{21} & \rho_{22} & \rho_{23} & \rho_{24} \\
\rho_{31} & \rho_{32} & \rho_{33} & \rho_{34} \\
\rho_{41} & \rho_{42} & \rho_{43} & \rho_{44}
\end{array}\right]
\end{eqnarray}
one can associate the realigned matrix
\begin{eqnarray}
\rho^\rea = \left[\begin{array}{cccc}
\rho_{11} & \rho_{12} & \rho_{21} & \rho_{22} \\
\rho_{13} & \rho_{14} & \rho_{23} & \rho_{24} \\
\rho_{31} & \rho_{32} & \rho_{41} & \rho_{42} \\
\rho_{33} & \rho_{34} & \rho_{43} & \rho_{44}
\end{array}\right].
\end{eqnarray}

The next step is to observe that there is a precise link between the
singular values of the realigned matrix $\rho^\rea$ and the Schmidt
coefficients of the density operator $\hat{\rho}$. Indeed, let us
denote by $\hhilb$, $\hhilba$ and $\hhilbb$ the Hilbert spaces of
linear operators in $\mathcal{H}$, $\hilba$ and $\hilbb$,
respectively, endowed with the Hilbert-Schmidt (HS) scalar product
(we are dealing with finite-dimensional vector spaces). Then, we
have that
\begin{equation} \label{tpdec}
\hhilb=\hhilba\otimes\hhilbb,
\end{equation}
and one can consider the Schmidt decomposition of a density operator
$\hat{\rho}\in\hhilb$ with respect to the tensor product
decomposition~{(\ref{tpdec})}. To this aim, let us fix a product
(orthonormal) basis in the Hilbert space
$\hhilb=\hhilba\otimes\hhilbb$. It will be convenient to choose the
basis formed by the partial isometries:
\begin{equation} \label{basis}
\bas_{(m n)(\mu\nu)}^\phant\equiv (|m\rangle\langle n|)\otimes
(|\mu\rangle\langle\nu|),\ \ \ |m\rangle\langle n|\in\hhilba,\
|\mu\rangle\langle\nu|\in\hhilbb,
\end{equation}
with $m,n=1,\ldots,\na$, $\mu,\nu=1,\ldots,\nb$. At this point, we
can write:
\begin{equation} \label{expa}
\hat{\rho}=\sum_{m,n,\mu,\nu} c_{(mn)(\mu\nu)}^\phant\, \bas_{(m
n)(\mu\nu)}^\phant,
\end{equation}
where the coefficients of the expansion~{(\ref{expa})} are given by
\begin{eqnarray}
c_{(mn)(\mu\nu)}^\phant   =  \left\langle \bas_{(m
n)(\mu\nu)}^\phant, \hat{\rho}\right\rangle_\hhilb & = & \tr
\!\left(\bas_{(mn)(\mu\nu)}^{\,\dagger}\hspace{0.3mm}
\hat{\rho}\right)
\nonumber \\
& = & \tr \!\left(\bas_{(nm)(\nu\mu)}^\phant\hspace{0.3mm}
\hat{\rho}\right)
\nonumber \\
& = & \langle m|\langle\mu
|\hspace{0.4mm}\hat{\rho}\hspace{0.4mm}|n\rangle|\nu\rangle\equiv
\rho_{(m\mu)(n\nu)}^\pha.
\end{eqnarray}
Hence, recalling relation~{(\ref{relrea})}, we conclude that the
matrix of coefficients $\left[c_{(mn)(\mu\nu)}^\phant\right]$
coincides with the realigned matrix $\rho^\rea$. We have thus
obtained the following:
\begin{proposition} \label{svrm} For any
density operator $\hat{\rho}\in\hhilb$, the Schmidt coefficients of
$\hat{\rho}$, with respect to the tensor product decomposition
$\hhilba\otimes\hhilbb$ of $\hhilb$, coincide with the singular
values of the associated realigned matrix $\rho^\rea$.
\end{proposition}
It this clear that this result extends to all operators in
$\mathcal{H}$: \emph{for any linear operator $\hat{A}$ in
$\hilba\otimes\hilbb$
--- regarded as a vector of $\hhilba\otimes\hhilbb$ --- the Schmidt coefficients of
$\hat{A}$ coincide with the singular values of the realigned matrix
$A^\rea$}. This fact will be used in Sect.~{\ref{new}}.

The last step is to extend the characterization of separable pure
states in terms of the singular values of the associated realigned
matrices to all states. Using Proposition~{\ref{svrm}}, one easily
gets to the RC. The RC provides a necessary condition for the
separability of bipartite quantum states, which is known to be
independent of the PPT criterion~\cite{ReCr}. The RC relies on the
evaluation of the trace norm of the realigned matrix $\rho^\rea$,
which is equal to the sum of the singular values of $\rho^\rea$. We
stress that, if $\hat{\rho}$ is a \emph{mixed} (i.e.\ non-pure)
state, then the SVD $\rho^\rea = \mathcal{U}\, \Lambda\,
\mathcal{V}$, with singular values $\{ \lambda_k
\}_{k=1,\dots,\dd^2}$, does not present the simple Kronecker product
structure as in~{(\ref{pure_SVD})}.

In order to obtain the RC, we argue as follows. Denoted by
$\|\cdot\|_{\tr}$ the trace norm, we have:
\begin{equation} \label{tracenorm}
\|\rho^\rea\|_{\tr} := \tr(|\rho^\rea|) =
\tr\!\left(((\rho^\rea)^\dag
\rho^\rea)^{\frac{1}{2}}\right)=\sum_{k=1}^{\dd^2} \lambda_k \, .
\end{equation}
Now, in the special case of a simply separable state
$\hrho=\hrho^\sista\otimes\hrho^\sistb$, we have that the Schmidt
coefficients of $\hrho$ (equivalently, the singular values of
$\rho^\rea$) are given by:
\begin{equation}
(\lambda_1=\|\hrho^\sista\otimes\hrho^\sistb\|_{\hhilb},\lambda_2=0,\ldots,\lambda_{\dd^2}=0),
\end{equation}
where we denote by $\|\cdot\|_{\hhilb}$ (alternatively, by
$\|\cdot\|_{\hhilba}$ and $\|\cdot\|_{\hhilbb}$) the HS norm in
$\hhilb$ (respectively, in $\hhilba$ and $\hhilbb$). Observe that
the following relation holds:
\begin{equation}
\|\hrho^\sista\otimes\hrho^\sistb\|_{\hhilb}=\|\hrho^\sista\|_{\hhilba}\,
\|\hrho^\sistb\|_{\hhilbb}=
\sqrt{\tr((\hrho^\sista)^2)}\,\sqrt{\tr((\hrho^\sistb)^2)};
\end{equation}
hence:
\begin{equation}
\|\hrho^\sista\otimes\hrho^\sistb\|_{\hhilb}\le 1,\ \mbox{and} \ \ \
\|\hrho^\sista\otimes\hrho^\sistb\|_{\hhilb}= 1\, \Leftrightarrow\,
\hrho^\sista,\hrho^\sistb\; \mbox{are pure states}.
\end{equation}
Therefore, the Schmidt coefficients of $\hrho$ are characterized by
the relations
\begin{equation}
\lambda_1\le 1,\ \lambda_2=\ldots=\lambda_{\dd^2}=0,\ \mbox{and} \ \
\ \lambda_1= 1\, \Leftrightarrow\, \hrho^\sista,\hrho^\sistb\;
\mbox{are pure states}.
\end{equation}
From these relations, recalling formula~{(\ref{tracenorm})}, we get
the following:
\begin{proposition} \label{simplysep}
If $\hrho\in\hhilba\otimes\hhilbb$ is a simply separable state ---
$\hrho=\hrho^\sista\otimes\hrho^\sistb$, $\hrho^\sista\in\hhilba$,
$\hrho^\sistb\in\hhilbb$
--- then the associated realigned matrix $\rho^\rea$ satisfies:
\begin{equation} \label{ineq}
\|\rho^\rea\|_{\tr} = \sum_{k=1}^{\dd^2} \lambda_k=\lambda_1\le 1 .
\end{equation}
Inequality~{(\ref{ineq})} is saturated if and only if the simply
separable state $\hrho$ is pure:
\begin{equation}
\|\rho^\rea\|_{\tr} = 1\ \Leftrightarrow \ \hrho=\hrho^2.
\end{equation}
\end{proposition}
Since separable states are convex superpositions of simply separable
states, we can exploit Proposition~{\ref{simplysep}} in order to
obtain a separability criterion. To this aim, it will be convenient
to introduce the following notations. Denoted by
$\mathcal{M}(N_1,N_2)$ the vector space of $N_1\times N_2$
complex-valued matrices, we define the \emph{linear application}
\begin{equation}
\rema :\, \hhilb \rightarrow
\mathcal{M}(N_{\sistaa}^2,N_{\sistbb}^2),\ \hat{A}\mapsto A^\rea.
\end{equation}
At this point, given a density operator $\hrho\in\hhilb$
--- separable with respect to the tensor product decomposition
$\mathcal{H}=\hilba \otimes \hilbb$, i.e.
\begin{equation}
\hrho = \sum_i p_i\; \hrho_i^\sista \otimes \hrho_i^\sistb,\ \ \
\hrho_i^\sista\in\hhilba,\ \hrho_i^\sistb\in\hhilbb,\ \ \
\mbox{(\emph{finite} convex superposition)}
\end{equation}
with $p_i > 0$ and $\sum_i p_i = 1$ --- we can apply
inequality~{(\ref{ineq})} (for a simply separable state) and the
triangle inequality:
\begin{eqnarray}
\|\rema (\hrho)\|_{\tr} & = & \|\rema (\mbox{$\sum_i p_i$}\;
\hrho_i^\sista \otimes \hrho_i^\sistb)\|_{\tr}
\nonumber\\
& = & \| \mbox{$\sum_i p_i$}\, \rema (\hrho_i^\sista \otimes
\hrho_i^\sistb)\|_{\tr} \ \ \ \ \ \ \mbox{(linearity of $\rema$)}
\nonumber\\
& \le & \mbox{$\sum_i p_i$}\, \| \rema (\hrho_i^\sista \otimes
\hrho_i^\sistb)\|_{\tr} \ \ \ \ \ \ \mbox{(triangle inequality)}
\nonumber\\
& \le & \mbox{$\sum_i p_i$} = 1. \hspace{2.65cm}
\mbox{(inequality~{(\ref{ineq})})}
\end{eqnarray}
We have thus obtained the following result:
\begin{theorem}[`realignment criterion'] Let $\hrho$ be a state in
$\mathcal{H}=\hilba\otimes\hilbb$. If $\hrho$ is separable, then the
associated realigned matrix $\rho^\rea$ (relative to any product
orthonormal basis in $\hilba\otimes\hilbb$) satisfies:
\begin{equation} \label{ine}
\|\rho^\rea\|_{\tr} = \sum_{k=1}^{\dd^2} \lambda_k\le 1 .
\end{equation}
Inequality~{(\ref{ine})} is saturated if the separable state $\hrho$
is pure.
\end{theorem}

\section{A class of inequalities inducing new separability criteria}
\label{new}

In the present section, we will derive a class of inequalities
satisfied by all separable states of a bipartite quantum system. As
in the case of the RC, these inequalities can be exploited for
detecting entanglement. We will use arguments similar to the one
adopted for deriving the standard RC, and we will keep the notations
and the assumptions introduced in the preceding section. In
particular, for any linear operator $\hat{A}$ in $\mathcal{H}$, we
will denote by $\rema(\hat{A})$ the realigned matrix associated with
$\hat{A}$, relatively to an arbitrarily fixed product orthonormal
basis in $\mathcal{H}$ (recall also that $\rema :\, \hhilb
\rightarrow \mathcal{M}(N_{\sistaa}^2,N_{\sistbb}^2)$ is a linear
map). In view of the separability criteria introduced below, we
stress also that the (easily computable) positive number $\|\rema
(\hat{A})\|_{\tr}$ is equal to the trace norm of the matrix of
coefficients of $\hat{A}$ --- regarded as a vector of
$\hhilba\otimes\hhilbb$ --- with respect to \emph{any} product
orthonormal basis in $\hhilba\otimes\hhilbb$ (as we have seen,
$\rema (\hat{A})$ is the matrix of coefficients associated with a
basis of the special type~{(\ref{basis})}). Moreover, it is also
worth mentioning the fact that the map
\begin{equation}
\|\rema(\cdot)\|_{\tr}:\hhilba\otimes\hhilbb\rightarrow\mathbbm{R}^+
\end{equation}
is a norm (actually, it is a \emph{cross norm} on
$\hhilba\otimes\hhilbb$; see~{\cite{CCN}}). In the following, we
will denote by $\dens$ ($\densa$, $\densb$) the convex subset of
$\hhilb$ (respectively, of $\hhilba$, $\hhilbb$) consisting of all
density operators in $\mathcal{H}$ (respectively, in $\hilba$,
$\hilbb$).

Let $\hrho\in\dens$ be a separable state in the bipartite Hilbert
space $\mathcal{H}=\hilba\otimes\hilbb$, with a separability
decomposition of the form
\begin{equation} \label{super}
\hrho = \sum_i p_i\; \hrho_i^\sista \otimes \hrho_i^\sistb,\ \ \
\mbox{(\emph{finite} convex superposition)}
\end{equation}
where:
\begin{equation}
\sum_{i} p_i = 1,\ \ \mbox{and}\ \ p_i>0,\ \forall i.
\end{equation}
Let us denote by $\mara$, $\marb$ the \emph{marginals} (reduced
density operators) of $\hrho$, namely:
\begin{equation}
\mara:=\trb(\hrho)=\sum_i p_i\, \hrho_i^\sista,\ \ \
\marb:=\tra(\hrho)=\sum_i p_i\, \hrho_i^\sistb.
\end{equation}

Now, given $2\mathsf{n}$ linear or antilinear (super-)operators
\begin{equation}
\hspace{-8mm} \eau :\hhilba\rightarrow\hhilba,\, \ldots, \ean
:\hhilba\rightarrow\hhilba,\ \; \ebu :\hhilbb\rightarrow\hhilbb,\,
\ldots, \ebn :\hhilbb\rightarrow\hhilbb,
\end{equation}
with $\mathsf{n}\ge 1$ --- precisely: we require these operators to
be \emph{jointly} linear or antilinear, i.e.\ either all linear or
all antilinear, in such a way that one can consistently define
tensor products and sums
--- we will associate with $\hrho$ the linear operator $\op :
\mathcal{H}\rightarrow\mathcal{H}$ defined by
\begin{eqnarray}
\op & := & \mathsf{n}^{-1} \left(\eau\otimes\ebu +\cdots +
\ean\otimes\ebn\right)(\hrho)
\nonumber \\ \label{defop}
& + &
\mathsf{n}^{-1} \Big(\sum_{k\neq
l}\eak\otimes\ebl\Big)(\mara\otimes\marb).
\end{eqnarray}
Of course, $\op$ will not be, in general, a density operator. Notice
that the operator $\op$ is defined in terms of the explicitly known
operators $\hrho$, $\mara$ and $\marb$; i.e.\
definition~{(\ref{defop})} does not involve the (in general)
`unknown density operators' that appear in the r.h.s.\ of the
separability decomposition~{(\ref{super})}. It is easy to check,
however, that the following relation holds:
\begin{eqnarray}
\hspace{-1.4cm} \op  &  = &
\frac{1}{\mathsf{n}}\sum_{i_1,\ldots,\indn} p_{i_1}\cdots
p_{\indn}\,
(\eau(\hrho_{i_1}^\sista)+\cdots+\ean(\hrho_{\indn}^\sista))\otimes
(\ebu(\hrho_{i_1}^\sistb)+\cdots+\ebn(\hrho_{\indn}^\sistb))
\nonumber\\ \label{fonda} &  \equiv &
\frac{1}{\mathsf{n}}\sum_{i_1,\ldots,\indn} p_{i_1}\cdots
p_{\indn}\,
(\eau(\hrho_{i_1}^\sista)+\cdots+\ean(\hrho_{\indn}^\sista)) \otimes
\abi ,
\end{eqnarray}
where the symbol $\abi$ denotes repetition of the preceding term
with the substitution of the subsystem $\mathsf{A}$ with the
subsystem $\mathsf{B}$ (all the rest remaining unchanged). For
instance, in order to check relation~{(\ref{fonda})}, consider that:
\begin{eqnarray}
\sum_{i_1,\ldots,\indn} p_{i_1}\cdots p_{\indn}\,
\eau(\hrho_{i_1}^\sista)\otimes\ebu(\hrho_{i_1}^\sistb)& = &
\sum_{i_1}p_{i_1}
\left(\eau\otimes\ebu\right)(\hrho_{i_1}^\sista\otimes\hrho_{i_1}^\sistb)
\nonumber\\
& = & \left(\eau\otimes\ebu\right)(\hrho),
\end{eqnarray}
where, since $p_i\in\mathbb{R}$, both linearity and antilinearity
work without distinction. In particular, for $\mathsf{n}=1$, we have
that $\opu:=\left(\eau\otimes\ebu\right)(\hrho)=\sum_{i}p_{i}
\left(\eau\otimes\ebu\right)(\hrho_{i}^\sista\otimes\hrho_{i}^\sistb)$,
and, for $\mathsf{n}=2$, we have:\footnote{In the special case where
the operators $\eau,\ebu$ and $\eadu,\ebdu$ coincide with the
identity and $-1$ times the identity, respectively, the operator
$\ope$ reduces to the expression $\hrho-\mara\otimes\marb$ which is
central in the separability criterion obtained in
ref.~{\cite{Zhang}}; i.e.: $\|\rema (\hrho -
\mara\otimes\marb)\|_{\tr} \le\sqrt{\left(1-\tr(\mara^2)\right)
\left(1-\tr(\marb^2)\right)}$, for every separable state
$\hrho\in\dens$. See also Corollary~{\ref{coro2}} below. }
\begin{equation}\hspace{-1.4cm}
\ope  :=  \frac{1}{2} \left(\eau\otimes\ebu +
\eadu\otimes\ebdu\right)(\hrho)
 + \frac{1}{2} \left(\eau\otimes\ebdu +
\eadu\otimes\ebu\right)(\mara\otimes\marb),
\end{equation}
and
\begin{eqnarray}
\ope & = & \frac{1}{2}\sum_{i,j} p_i p_j\,
(\eau(\hrho_i^\sista)+\eadu(\hrho_j^\sista))\otimes(\ebu(\hrho_i^\sistb)+\ebdu(\hrho_j^\sistb))
\nonumber\\
& \equiv & \frac{1}{2}\sum_{i,j} p_i p_j\,
(\eau(\hrho_i^\sista)+\eadu(\hrho_j^\sista))\otimes \abi .
\end{eqnarray}

At this point, in order to obtain the new class of inequalities
announced above, we argue as follows. First of all, just for the
sake of notational conciseness, we will consider in our derivation
the operator $\ope$ rather than the more general expression $\op$
(the extension of our argument to $\op$ is straightforward). Then,
we observe that
\begin{eqnarray}
\hspace{-8mm}\|\rema (\ope)\|_{\tr} & = & \frac{1}{2}\,\|\rema
(\mbox{$\sum_{i,j}$}\, p_i p_j\,
(\eau(\hrho_i^\sista)+\eadu(\hrho_j^\sista))\otimes\abi)\|_{\tr}
\nonumber\\
& = & \frac{1}{2}\,\|\mbox{$\sum_{i,j}$}\, p_i p_j\,\rema (
(\eau(\hrho_i^\sista)+\eadu(\hrho_j^\sista))\otimes\abi)\|_{\tr}
\nonumber\\ \label{diseg1}
& \le &
\frac{1}{2}\,\mbox{$\sum_{i,j}$}\,p_i p_j\,\| \rema (
(\eau(\hrho_i^\sista)+\eadu(\hrho_j^\sista))\otimes\abi)\|_{\tr},
\end{eqnarray}
where for obtaining the last line we have used the triangle
inequality. Next, consider that the Schmidt coefficients of the
operator
$(\eau(\hrho_i^\sista)+\eadu(\hrho_j^\sista))\otimes\abi\in\hhilb$
are given by
\begin{equation}
(\lambda_1=\|(\eau(\hrho_i^\sista)+\eadu(\hrho_j^\sista))\otimes\abi\|_{\hhilb},
\lambda_2=0,\ldots,\lambda_{\dd^2}=0),
\end{equation}
since the Schmidt decomposition of this operator consists of a
single term. Hence, we have:
\begin{eqnarray}
\hspace{-2cm}
\|\rema((\eau(\hrho_i^\sista)+\eadu(\hrho_j^\sista))\otimes
\abi)\|_{\tr} & = & \|\eau(\hrho_i^\sista) +
\eadu(\hrho_j^\sista)\|_{\hhilba}\, \|\ebu(\hrho_i^\sistb) +
\ebdu(\hrho_j^\sistb)\|_{\hhilbb}
\nonumber\\
& = & \Big(\langle
\eau(\hrho_i^\sista),\eau(\hrho_i^\sista)\rangle_{\hhilba} +
\langle\eadu(\hrho_j^\sista),\eadu(\hrho_j^\sista)\rangle_{\hhilba}
\nonumber\\
& + & (\langle
\eau(\hrho_i^\sista),\eadu(\hrho_j^\sista)\rangle_{\hhilba}
+\coco)\Big)^{\frac{1}{2}} \abi^{\frac{1}{2}} ,
\end{eqnarray}
where $\langle\cdot,\cdot\rangle_{\hhilba}$ is the (Hilbert-Schmidt)
scalar product in $\hhilba$. Thus, if we assume that, for some
$\supa,\supb \ge 0$,
\begin{equation}
\hspace{-1cm}\|\eau(\hsi_1^\sista)\|_{\hhilba}^2+
\|\eadu(\hsi_2^\sista)\|_{\hhilba}^2\le 2\supa,\ \ \;
\|\ebu(\hsi_1^\sistb)\|_{\hhilbb}^2 +
\|\ebdu(\hsi_2^\sistb)\|_{\hhilbb}^2\le 2\supb,
\end{equation}
$\forall\, \hsi_1^\sista,\hsi_2^\sista\in\densa$, $\forall\,
\hsi_1^\sistb, \hsi_2^\sistb\in\densb$, we find the following
estimate:
\begin{eqnarray}
\hspace{-1.8cm}\frac{
\|\rema((\eau(\hrho_i^\sista)+\eadu(\hrho_j^\sista))\otimes
\abi)\|_{\tr}}{2} & \le & \sqrt{\left(\supa+\frac{1}{2}\Big(\langle
\eau(\hrho_i^\sista),\eadu(\hrho_j^\sista)\rangle_{\hhilba}
+\coco\Big)\right)} \nonumber\\ \label{diseg2} & \times &
\sqrt{\abi}.
\end{eqnarray}
Eventually, from inequalities~{(\ref{diseg1})} and~{(\ref{diseg2})},
we get:
\begin{eqnarray}
\|\rema (\ope)\|_{\tr} & \le & \mbox{$\sum_{i,j}$}\,\sqrt{p_i p_j
\left(\supa+\frac{1}{2}\Big(\langle
\eau(\hrho_i^\sista),\eadu(\hrho_j^\sista)\rangle_{\hhilba}
+\coco\Big)\right)}
\nonumber\\
& \times &    \sqrt{p_i p_j \left(\supb +\frac{1}{2}\Big(\langle
\ebu(\hrho_i^\sistb),\ebdu(\hrho_j^\sistb)\rangle_{\hhilbb}
+\coco\Big)\right)}
\nonumber\\
& \le & \sqrt{\left(\supa+\frac{1}{2}\Big(\mbox{$\sum_{i,j}$}\,p_i
p_j\,\langle
\eau(\hrho_i^\sista),\eadu(\hrho_j^\sista)\rangle_{\hhilba}
+\coco\Big)\right)}
\nonumber\\
& \times & \sqrt{\abi} ,
\end{eqnarray}
where the second inequality above follows from the Cauchy-Schwarz
inequality. Therefore, for every separable state $\hrho\in\dens$ we
have:
\begin{equation} \label{checkine}
\hspace{-1cm}\|\rema (\ope)\|_{\tr} \le
\sqrt{\left(\supa+\frac{1}{2}\Big(\langle
\eau(\mara),\eadu(\mara)\rangle_{\hhilba} +\coco\Big)\right)} \
\sqrt{\abi}.
\end{equation}

We notice that, if $\vhrho$ is a \emph{generic} state in $\dens$
(i.e., separable or not), then one can define an operator $\vope$
precisely as it has been done for the operator $\ope$ associated
with a separable state $\hrho$. Moreover, observe that we have:
\begin{eqnarray}
\hspace{-2cm}0\le\frac{ \|\rema((\eau(\vmara)+\eadu(\vmara))\otimes
\abi)\|_{\tr}^2}{4} & \le & \left(\supa+\frac{1}{2}\Big(\langle
\eau(\vmara),\eadu(\vmara)\rangle_{\hhilba} +\coco\Big)\right)
\nonumber\\ \label{diseg3} & \times & \abi,
\end{eqnarray}
where $\vmara$, $\vmarb$ are the marginals of $\vhrho$. Hence, it
makes sense to check inequality~{(\ref{checkine})} for a
\emph{generic} state (the square roots on the r.h.s.\ of
inequality~{(\ref{checkine})} are well defined). This inequality may
not be satisfied by some state which can be then detected as an
entangled state.

Extending the above proof to the operator $\op$, $\mathsf{n}\ge 1$,
associated with a separable state $\hrho\in\dens$, we obtain the
following result:
\begin{theorem}[`generalized RC'] \label{nuovo-nuovo}
Let $\hrho$ be a state in $\mathcal{H}=\hilba\otimes\hilbb$ and
\begin{equation} \label{chosen}
\hspace{-6mm} \eau :\hhilba\rightarrow\hhilba,\, \ldots, \ean
:\hhilba\rightarrow\hhilba,\ \; \ebu :\hhilbb\rightarrow\hhilbb,\,
\ldots, \ebn :\hhilbb\rightarrow\hhilbb,
\end{equation}
$\mathsf{n}\ge 1$, jointly linear or antilinear operators such that,
for some $\supa,\supb \ge 0$,
\begin{equation} \label{condicio}
\hspace{-2.3cm} \|\eau(\hsi_1^\sista)\|_{\hhilba}^2+\cdots+
\|\ean(\hsi_{\ennebis}^\sista)\|_{\hhilba}^2\le \mathsf{n}\supa,\ \
\; \|\ebu(\hsi_1^\sistb)\|_{\hhilbb}^2 +\cdots+
\|\ebn(\hsi_{\ennebis}^\sistb)\|_{\hhilbb}^2\le \mathsf{n}\supb,
\end{equation}
$\forall\, \hsi_1^\sista,\ldots ,\hsi_{\ennebis}^\sista\in\densa$,
$\forall\, \hsi_1^\sistb,\ldots , \hsi_{\ennebis}^\sistb\in\densb$,
and consider the linear operator $\op$ in $\mathcal{H}$ defined by
\begin{equation}
\op :=  \mathsf{n}^{-1}
\Big(\sum_{k=1}^{\mathsf{n}}\eak\otimes\ebk\, (\hrho)+\sum_{k\neq
l}\eak\otimes\ebl\,(\mara\otimes\marb)\Big),
\end{equation}
where $\mara$ and $\marb$ are the marginals of $\hrho$, namely:
$\mara:=\trb(\hrho)$, $\marb:=\tra(\hrho)$. If the state $\hrho$ is
separable, then the following inequality holds:
\begin{equation} \label{inefond}
\hspace{-1cm} \|\rema (\op)\|_{\tr} \le
\sqrt{\Big(\supa+\frac{1}{\mathsf{n}}\sum_{k < l}\big(\langle
\eak(\mara),\eal(\mara)\rangle_{\hhilba} +\coco\big)\Big)\abi}.
\end{equation}
\end{theorem}
It is obvious that, once chosen the operators~{(\ref{chosen})},
inequality~{(\ref{inefond})} is satisfied, in particular, if we set
$\supa=\osupa$ and $\supb=\osupb$, where $\osupa$ is the positive
number defined by
\begin{equation} \label{defosupa}
\osupa:=\sup\Big\{\mathsf{n}^{-1}\sum_{k=1}^{\mathsf{n}}
\|\eak(\hsi_k^\sista)\|_{\hhilba}^2\colon\;  \hsi_1^\sista,\ldots
,\hsi_{\ennebis}^\sista\in\densa\Big\},
\end{equation}
and $\osupb$ is defined analogously.

The reason why we call Theorem~{\ref{nuovo-nuovo}} `generalized RC'
is clear: for $\mathsf{n}=1$, and $\eau,\ebu$ coinciding with the
identity (super-)operators (so that we can set: $\supa=\supb=1$), we
recover the RC; see also Corollary~{\ref{coro2}} below. We stress
that, since it makes sense (as already observed) to check
inequality~{(\ref{inefond})} for a generic state,
Theorem~{\ref{nuovo-nuovo}} induces, potentially, a whole new class
of separability criteria. In this regard, it is easy to see that if
one \emph{fixes a priori} in the r.h.s.\ of each of
inequalities~{(\ref{condicio})} arbitrary strictly positive numbers
$\supa$ and $\supb$ --- e.g., if one sets $\supa=\supb=1$
--- then the class of \emph{independent} separability criteria
induced by inequality~{(\ref{inefond})} is not restricted by such a
specific choice. A certain criterion --- i.e., a certain set of
operators of the type~{(\ref{chosen})} satisfying
inequalities~{(\ref{condicio})}, and such that
inequality~{(\ref{inefond})} is violated by some entangled state
--- will be \emph{optimal} if $\supa=\osupa$ and $\supb=\osupb$, with
$\osupa$, $\osupb$ defined as in~{(\ref{defosupa})}; otherwise, one
can obtain an optimal criterion by suitably rescaling the operators
(provided that one is able to evaluate the numbers $\osupa$ and
$\osupb$).

Notice that, since the HS norm of density operators is not larger
than one, condition~{(\ref{condicio})} is satisfied if for the norm
of the (super-)operators $\{\eak,\ebk\}_{k=1,\ldots,\mathsf{n}}$ the
following inequalities hold:
\begin{equation}
\sum_{k=1}^{\mathsf{n}}\|\eak\|^2\le \mathsf{n}\supa,\ \ \;
\sum_{k=1}^{\mathsf{n}}\|\ebk\|^2\le \mathsf{n}\supb;
\end{equation}
in particular, if the norm of the operators
$\{\eak,\ebk\}_{k=1,\ldots,\mathsf{n}}$ is not larger than
$\sqrt{\supa}$ or $\sqrt{\supb}$, respectively. For instance, one
can choose the operators $\{\eak,\ebk\}_{k=1,\ldots,\mathsf{n}}$ to
be either (all) unitary or (all) antiunitary in such a way that
inequality~{(\ref{inefond})} is satisfied with
$\supa=\supb=1$.\\
Observe that condition~{(\ref{condicio})} is also satisfied --- with
$\supa=\supb=1$ --- if the (super-)operators
$\{\eak,\ebk\}_{k=1,\ldots,\mathsf{n}}$ are such that
\begin{equation} \label{condicio2}
\|\eak(\hsi^\sista)\|_{\hhilba}\le 1,\ \
\|\ebk(\hsi^\sistb)\|_{\hhilbb}\le 1,\ \ k=1,\ldots,\mathsf{n},
\end{equation}
$\forall\, \hsi^\sista\in\densa$, $\forall\, \hsi^\sistb\in\densb$.
One can assume, in particular, that they are
trace-norm-nonincreasing on positive operators (since the
Hilbert-Schmidt norm is majorized by the trace norm); for instance,
positive trace-preserving linear maps.

Another natural choice consists in taking linear or antilinear
operators $\{\xak,\yak\}_{k=1,\ldots,\mathsf{n}}$ in $\hilba$ and
$\{\xbk,\ybk\}_{k=1,\ldots,\mathsf{n}}$ in $\hilbb$, and setting:
\begin{equation}
\hspace{-1.9cm} \eak : \hhilba\ni\aaa\mapsto \xak \aaa\hspace{0.3mm}
\yak\in\hhilba,\ \ \; \ebk : \hhilbb\ni\bbb\mapsto\xbk
\bbb\hspace{0.3mm} \ybk\in\hhilbb,\ \ \; k=1,\ldots,\mathsf{n}\,;
\end{equation}
precisely, we need to fix the following constraint: the operators
$\{\xak,\yak,\xbk,\ybk\}_{k=1,\ldots,\mathsf{n}}$ must be linear or
antilinear in such a way that the (super-)operators
$\{\eak,\ebk\}_{k=1,\ldots,\mathsf{n}}$ are (well defined as
operators in $\hhilba$, $\hhilbb$ and) jointly linear or antilinear.
Hence, the operators
$\{\xak,\yak,\xbk,\ybk\}_{k=1,\ldots,\mathsf{n}}$ must be jointly
linear or antilinear as well: i.e., either all linear (so that the
corresponding super-operators
$\{\eak,\ebk\}_{k=1,\ldots,\mathsf{n}}$ are linear) or all
antilinear (in this case, the super-operators
$\{\eak,\ebk\}_{k=1,\ldots,\mathsf{n}}$ are antilinear).

Suppose, then, that the operators
$\{\xak,\yak,\xbk,\ybk\}_{k=1,\ldots,\mathsf{n}}$ are jointly linear
or antilinear. In this case, taking into account the fact that (due
to a well known relation between the standard operator norm
$\|\cdot\|$ and the HS norm)
\begin{equation}
\hspace{-1.4cm}\| \xak \hsi^\sista \yak\|_{\hhilba}\le  \|\xak\|\,
\|\yak\|\,\|\hsi^\sista\|_{\hhilba} ,\ \ \| \xbk \hsi^\sistb
\ybk\|_{\hhilbb}\le  \|\xbk\|\, \|\ybk\|\,\|\hsi^\sistb\|_{\hhilbb},
\end{equation}
$k=1,\ldots,\mathsf{n}$,  for any $\hsi^\sista\in\densa$ and
$\hsi^\sistb\in\densb$
--- where: $\|\hsi^\sista\|_{\hhilba}\le\|\hsi^\sista\|_{\tr}=1$ and
$\|\hsi^\sistb\|_{\hhilbb}\le 1$ --- from
Theorem~{\ref{nuovo-nuovo}} we obtain the following result:
\begin{corollary} \label{coruno}
Let $\hrho$ be a state in $\mathcal{H}=\hilba\otimes\hilbb$ and
\begin{equation}
\hspace{-5mm} \xau,\ \yau,\ldots ,\xan,\ \yan:\,
\hilba\rightarrow\hilba,\ \ \ \xbu,\ \ybu,\ldots , \xbn,\ \ybn:\,
\hilbb\rightarrow\hilbb,
\end{equation}
$\mathsf{n}\ge 1$, jointly linear or antilinear operators such that
\begin{equation}
\sum_{k=1}^{\mathsf{n}}\|\xak\|^2\, \|\yak\|^2 \le
\mathsf{n}\supa, \ \ \; \sum_{k=1}^{\mathsf{n}}\|\xbk\|^2\,
\|\ybk\|^2 \le \mathsf{n}\supb ,
\end{equation}
for some $\supa,\supb >0$, and consider the linear operator $\oper$
in $\mathcal{H}$ defined by
\begin{eqnarray}
\oper & := &
\frac{1}{\mathsf{n}}\sum_{k=1}^{\mathsf{n}}(\xak\otimes\xbk)\,\hrho\,(\yak\otimes\ybk)
\nonumber\\
& + & \frac{1}{\mathsf{n}}\sum_{k\neq
l}(\xak\otimes\xbl)(\mara\otimes\marb)(\yak\otimes\ybl),
\end{eqnarray}
where $\mara$ and $\marb$ are the marginals of $\hrho$. If $\hrho$
is separable, then the following inequality holds:
\begin{eqnarray}
\hspace{-1.5cm} \|\rema (\oper)\|_{\tr} & \le &
\sqrt{\Big(\supa+\frac{1}{\mathsf{n}}\sum_{k<l}\big(\langle
\xak\mara\yak,\xal\mara\yal\rangle_{\hhilba} +\coco\big)\Big)}
\nonumber\\
&\times &
\sqrt{\Big(\supb+\frac{1}{\mathsf{n}}\sum_{k<l}\big(\langle
\xbk\marb\ybk,\xbl\marb\ybl\rangle_{\hhilbb} +\coco\big)\Big)}.
\end{eqnarray}
\end{corollary}

In particular, we can set $\mathsf{n}=2$, and the operators
$\{\xak,\yak,\xbk,\ybk\}_{k=1,2}$ can be chosen to be \emph{unitary}
(hence, we can set: $\supa=\supb=1$). For the specific choice
\begin{equation}
\xau=e^{\mathrm{i}\alpha_1}\,\ida,\ \
\xadu=e^{\mathrm{i}\alpha_2}\,\ida,\ \
\xbu=e^{\mathrm{i}\beta_1}\,\idb,\ \
\xbdu=e^{\mathrm{i}\beta_2}\,\idb,
\end{equation}
\begin{equation}
\yau=e^{\mathrm{i}\gamma_1}\,\ida,\ \
\yadu=e^{\mathrm{i}\gamma_2}\,\ida,\ \
\ybu=e^{\mathrm{i}\delta_1}\,\idb,\ \
\ybdu=e^{\mathrm{i}\delta_2}\,\idb,
\end{equation}
--- setting
\begin{equation}
\alpha_1+\beta_1+\gamma_1+\delta_1=\omega
\end{equation}
and
\begin{equation}
-\alpha_1 +\alpha_2-\gamma_1+\gamma_2=\theta,\ \ \ -\beta_1
+\beta_2-\delta_1+\delta_2=\phi,
\end{equation}
so that we have:
\begin{equation}
\alpha_2+\beta_2+\gamma_2+\delta_2=\omega+\theta+\phi
\end{equation}
and
\begin{equation}
\alpha_1 +\beta_2+\gamma_1+\delta_2=\omega+\phi,\ \ \ \alpha_2
+\beta_1+\gamma_2+\delta_1=\omega+\theta
\end{equation}
---  we find the following result:
\begin{corollary} \label{coro2}
Let $\hrho$ be a state in $\mathcal{H}=\hilba\otimes\hilbb$. If
$\hrho$ is separable, then the following inequality holds:
\begin{eqnarray} \label{ine-ine}
\hspace{-2.4cm} \left\|\rema
\!\left(\left(\frac{e^{\mathrm{i}\omega}+e^{\mathrm{i}(\omega+\theta+\phi)}}{2}\right)\hrho
+\!\left(\frac{e^{\mathrm{i}(\omega+\theta)}+e^{\mathrm{i}(\omega+\phi)}}{2}\right)
\mara\otimes\marb\right)\right\|_{\tr} & \le &
\sqrt{1+\cos\theta\,\tr(\mara^2)} \nonumber\\
&\times & \sqrt{1+\cos\phi\,\tr(\marb^2)},
\end{eqnarray}
for any $\omega,\theta,\phi\in \mathbbm{R}$, where $\mara$ and
$\marb$ are the marginals of $\hrho$. In particular, for $\omega=0$
and $\phi=-\theta$, we have that
\begin{equation} \label{thetaine}
\|\rema (\hrho +\cos\theta\, \mara\otimes\marb)\|_{\tr} \le
\sqrt{\left(1+\cos\theta\,\tr(\mara^2)\right)
\left(1+\cos\theta\,\tr(\marb^2)\right)},
\end{equation}
for any $\theta\in [0,\pi]$, while, for $\omega=0$ and
$\phi=\theta$, we have that
\begin{eqnarray} \label{ine-ine-ine}
\hspace{-2cm} \left\|\rema
\!\left(\left(\frac{1+e^{\mathrm{i}2\theta}}{2}\right)\hrho +
e^{\mathrm{i}\theta} \mara\otimes\marb\right)\right\|_{\tr} & \le &
\sqrt{\left(1+\cos\theta\,\tr(\mara^2)\right)
\left(1+\cos\theta\,\tr(\marb^2)\right)},
\end{eqnarray}
for any $\theta\in [0,2\pi[$.
\end{corollary}
We remark that the presence of the parameter $\omega$ on the l.h.s.\
of inequality~{(\ref{ine-ine})} is trivial. However, it is
convenient for deriving specific inequalities; for instance, setting
$\omega=-\theta=-\phi$, we get:
\begin{equation} \label{weget}
\hspace{-2cm} \|\rema (\cos\theta\,\hrho +
\mara\otimes\marb)\|_{\tr} \le
\sqrt{\left(1+\cos\theta\,\tr(\mara^2)\right)
\left(1+\cos\theta\,\tr(\marb^2)\right)},\ \ \theta\in [0,\pi].
\end{equation}

Observe that, fixing the value $\theta=\pi/2$ in the family of
inequalities~{(\ref{thetaine})}, we find again the standard RC. For
$\theta=\pi$, we recover the criterion derived in
ref.~{\cite{Zhang}}, where it is also shown that this separability
criterion is actually stronger than the standard RC. Notice that the
same criterion can also be obtained, for instance, from the family
of inequalities~{(\ref{ine-ine-ine})}, or~{(\ref{weget})}, setting
again $\theta=\pi$.

\section{Examples of entanglement detection}
\label{examples}

As already observed, Theorem~{\ref{nuovo-nuovo} allows to obtain a
new class of separability criteria for bipartite quantum systems.
This class seems to be very large and a wide range exploration of
the entanglement detection by means of such criteria is beyond the
scope of the present contribution. However, for the sake of
illustration, we will consider some simple example of application of
our results.

In particular, the subclass of inequalities~{(\ref{thetaine})}
induces a simple subclass of criteria characterized by the parameter
$\theta\in [0,\pi]$. Varying this parameter, one obtains a
`continuous family of separability criteria', which includes, as
already observed, the RC ($\theta=\pi/2$) and, for $\theta=\pi$, a
criterion recently proposed in ref.~{\cite{Zhang}}. Given a certain
class of states, one can try to detect entanglement applying these
criteria for different values of the parameter $\theta$.

As an example, we have considered a one-parameter family of
two-qutrit bound entangled states $\{\hrho(a)\}_{a\in [0,1]}$
presented in ref.~{\cite{Horo}}. In a given local basis for the
two-qutrit system, this family of states have the following matrix
representation:
\begin{equation}
\rho(a) = \frac{1}{8a+1} \left[\begin{array}{ccccccccc}
a & 0 & 0 & 0 & a & 0 & 0 & 0 & a \\
0 & a & 0 & 0 & 0 & 0 & 0 & 0 & 0 \\
0 & 0 & a & 0 & 0 & 0 & 0 & 0 & 0 \\
0 & 0 & 0 & a & 0 & 0 & 0 & 0 & 0 \\
a & 0 & 0 & 0 & a & 0 & 0 & 0 & a \\
0 & 0 & 0 & 0 & 0 & a & 0 & 0 & 0 \\
0 & 0 & 0 & 0 & 0 & 0 & \frac{1+a}{2} & 0 & \frac{\sqrt{1-a^2}}{2} \\
0 & 0 & 0 & 0 & 0 & 0 & 0 & a & 0 \\
a & 0 & 0 & 0 & a & 0 & \frac{\sqrt{1-a^2}}{2} & 0 & \frac{1+a}{2}
\end{array}\right],
\end{equation}
$0\le a\le 1$. We have then considered a statistical mixture of this
class of states with the maximally entangled state, hence obtaining
the two-parameter family of states:
\begin{equation} \label{thefamily}
\hrho(a,p)=p\,\hrho(a)+\frac{1}{9}(1-p)\,\ide\;,\ \ \ 0\le a\le 1,\
0\le p\le 1\,.
\end{equation}

\begin{figure}
\includegraphics[height=0.4\textwidth]{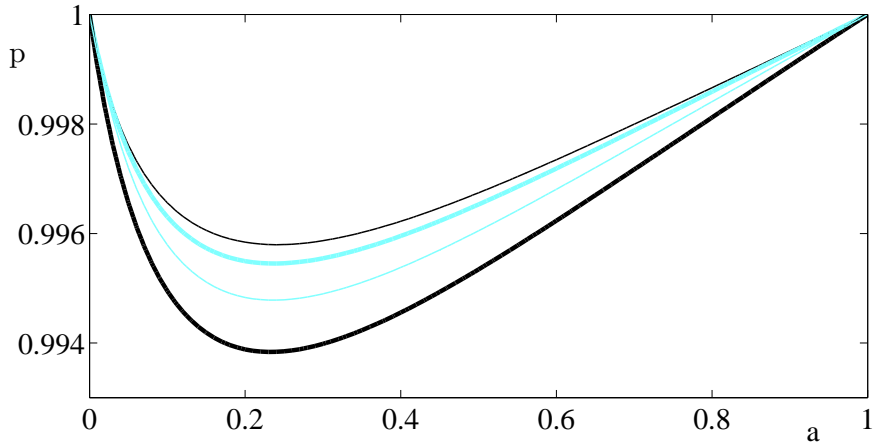} \caption{(color
online.) The curves in the plot identify those elements of the
family of states defined in~(\ref{thefamily}) for which
inequality~{(\ref{thetaine})} is saturated, in correspondence with
various values of the parameter $\theta$. The thin black line
corresponds to $\theta=0$, the thick light blue line to
$\theta=\pi/2$ (RC), the thin light blue line to $\theta=3\pi/4$,
and the thick black line to $\theta=\pi$. The inequalities are
violated by the (entangled) states associated with the values of the
parameters $a,p$ lying in the region above the curves. The criterion
corresponding to $\theta=\pi$ turns out to be the strongest in
detecting entangled states in the class of states considered.}
\label{theta}
\end{figure}

We have checked numerically inequalities~{(\ref{thetaine})} on this
set of states. As shown in figure~\ref{theta}, depending on the
value of the parameter $\theta$, the new separability criteria can
be stronger than the RC. The numerical calculations also indicate
that, on the family of states that we have considered, the strongest
criterion for the detection of entanglement is the one corresponding
to $\theta=\pi$.

One can consider another simple subclass of inequalities, namely,
the class associated with the following set of super-operators:
\begin{equation}
\eau = e^{\mathrm{i}\theta}\,\calida,\ \ \ebu =
e^{-\mathrm{i}\theta}\,\caltrb,\ \ \eadu = \calida,\ \ \ebdu =
\calidb,\ \ \theta\in [0,\pi],
\end{equation}
where $\calida\colon\hhilba\rightarrow\hhilba$,
$\calidb\colon\hhilbb\rightarrow\hhilbb$ are the identity
super-operators and $\caltrb\colon\hhilbb\rightarrow\hhilbb$ is the
transposition associated with a given orthonormal basis in $\hilbb$
(recall that transposition, differently from taking the adjoint, is
a basis-dependent map), one obtains the following family of
inequalities:
\begin{equation} \label{T_thetaine}
\|\rema (\ope)\|_{\tr} \le
\sqrt{\left(1+\cos\theta\,\tr(\mara^2)\right)
\left(1+\cos\theta\,\tr(\marbt\,\hrho_\sistb^\phanto)\right)},
\end{equation}
with
\begin{equation}
\ope = \frac{1}{2}\left( \hrhotb\hspace{-0.6mm} + \hrho \right) +
\frac{1}{2}\left( e^{\mathrm{i}\theta}\hspace{0.3mm}
\hrho_\sista^\phanto\otimes\hrho_\sistb^\phanto +
e^{-\mathrm{i}\theta}\hspace{0.3mm}
\hrho_\sista^\phanto\otimes\marbt\right),
\end{equation}
where $\marbt$ is the `transposed operator' (i.e.\
$\marbt\equiv\caltrb(\hrho_\sistb^\phanto)$), and $\hrhotb$ is the
`partially transposed operator' (i.e.\
$\hrhotb\equiv\calida\otimes\caltrb(\hrho)$).

We have then considered the family of two-qubit states introduced in
the second of papers~{\cite{CCN}}. In a given local basis for the
two-qubit system, they are expressed by a matrix of the form:
\begin{eqnarray}\label{T_family}
\rho(t,s,r) = \frac{1}{2}\left[
\begin{array}{cccc}
1+r & 0 & 0   & t \\
0   & 0 & 0   & 0 \\
0   & 0 & s-r & 0 \\
t   & 0 & 0   & 1-s
\end{array}
\right].
\end{eqnarray}
For $r=s/2$, it is easy to show that these states are well defined
in a domain containing the interval: $t\in [0, 0.25]$, $s\in
[0,0.9]$. Moreover, the specified family of states are known to be
separable if and only if $t=0$. As an example, we have checked
inequalities~(\ref{thetaine}) and~(\ref{T_thetaine}) (in this case,
the transposition $\caltrb$ is the one associated with the given
local basis), for different values of the parameter $\theta$, on the
specified family of states; namely, for $r=s/2$ with $t,s$ belonging
to the specified range: $t\in [0, 0.25]$, $s\in [0,0.9]$. The
corresponding plots are shown in figure~\ref{both}.

\begin{figure}
\includegraphics[height=0.4\textwidth]{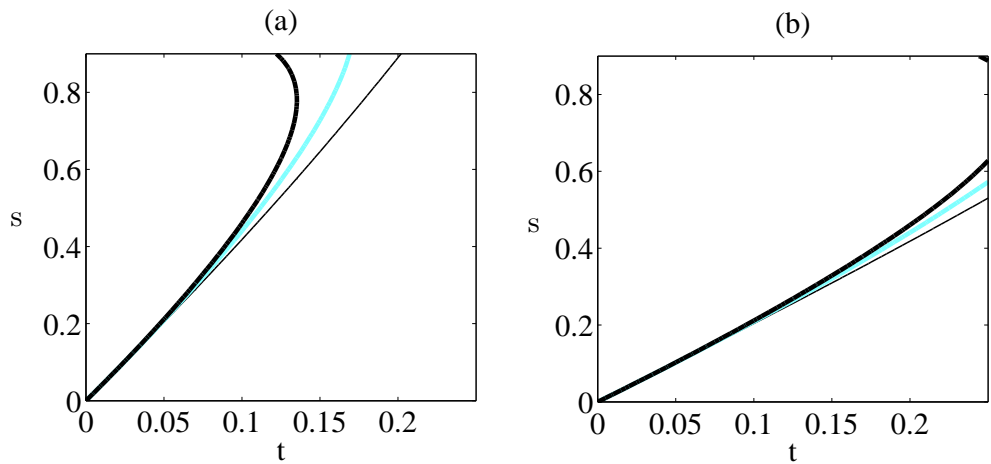}
\caption{(color online.) The curves in the plot identify those
elements of the family of states defined in~(\ref{T_family}) (with:
$r=s/2$, $t\in [0, 0.25]$, $s\in [0,0.9]$) for which
inequality~{(\ref{thetaine})} (diagram (a)) and
inequality~(\ref{T_thetaine}) (diagram (b)) are saturated, in
correspondence with three values of the parameter $\theta$. The thin
black line corresponds to $\theta=0$, the thick light blue line to
$\theta=3\pi/4$, and the thick black line to $\theta=\pi$. The
inequalities are violated by the (entangled) states associated with
the values of the parameters $t,s$ lying in the region on the right
with respect to the curves.} \label{both}
\end{figure}

\section{Conclusions} \label{conclusions}

In the present paper, we have introduced a class of inequalities for
bipartite quantum systems that are satisfied by separable states
and, hence, potentially induce new separability criteria. Each
inequality corresponds to a choice of suitable linear or antilinear
super-operators $\{\eak\}_{k=1,\ldots,\mathsf{n}}$ and
$\{\ebk\}_{k=1,\ldots,\mathsf{n}}$, respectively in the
Hilbert-Schmidt spaces $\hhilba$ and $\hhilbb$ associated with the
`local subsystems' $\mathsf{A}$ and $\mathsf{B}$ of the bipartite
quantum system; see Theorem~{\ref{nuovo-nuovo}}. A simple subclass
of inequalities are parametrized, in a natural way, by $\theta\in
[0,\pi]$; see inequality~{(\ref{thetaine})} in
Corollary~{\ref{coro2}}. This subclass contains, in particular, the
inequality at the base of the standard RC ($\theta=\pi/2$), and an
inequality ($\theta=\pi$) inducing a separability criterion which is
the main result obtained in ref.~{\cite{Zhang}}, where it is shown
that this criterion is actually stronger than the RC. We thus expect
the class of separability criteria induced by the inequalities
introduced here to be, in general, independent of the RC.

It is worth observing that another special subclass of inequalities
is obtained setting $\mathsf{n}=2$ and
\begin{equation}
\xau=\filta,\ \; \yau=(\filta)^\dagger,\ \; \xbu=\filtb,\ \;
\ybu=(\filtb)^\dagger,
\end{equation}
\begin{equation}
\xadu=e^{\mathrm{i}\theta}\,\filta,\ \; \yadu=(\filta)^\dagger,\ \;
\xbdu=e^{-\mathrm{i}\theta}\,\filtb,\ \; \ybdu=(\filtb)^\dagger,\ \;
\theta\in [0,\pi],
\end{equation}
--- where $\filta\colon\hilba\rightarrow\hilba$,
$\filtb\colon\hilbb\rightarrow\hilbb$ are linear operators such that
$\|\filta\|\le 1$, $\|\filtb\|\le 1$ (thus we can set
$\supa=\supb=1$)
--- in Corollary~{\ref{coruno}}. Hence, for every separable state
$\hrho\in \dens$, we have:
\begin{eqnarray}
\hspace{-1cm}\left\|\rema\hspace{-0.5mm}
\left(\filta\otimes\filtb\hspace{0.6mm}\hrho\hspace{1mm}(\filta)^\dagger\otimes
(\filtb)^\dagger+\hspace{0.3mm}\cos\theta\hspace{0.4mm}
(\filta\mara(\filta)^\dagger)\otimes(\filtb\marb(\filtb)^\dagger)\right)\right\|_{\tr}&
\le & \nonumber\\ \label{filters} \hspace{-1.4cm} \le
\sqrt{\left(1+\cos\theta\,\tr\hspace{-0.8mm}\left((\filta\mara(\filta)^\dagger)^2\right)\right)
\left(1+\cos\theta\,\tr\hspace{-0.8mm}\left((\filtb\marb(\filtb)^\dagger)^2\right)\right)}.&
&
\end{eqnarray}
For a suitable choice of the operators $\filta$, $\filtb$ (`local
filtering operations'), inequality~{(\ref{filters})} gives the
`local filtering enhancement' of the separability criterion induced
by inequality~{(\ref{thetaine})}. In particular, for $\theta=\pi/2$,
one obtains the `RC with local filtering'. In the case where
$\dim(\hilba)=\dim(\hilbb)$, this powerful separability criterion is
equivalent to the `covariance matrix criterion under filtering'.
See~{\cite{Hyllus}} and references therein. For $\theta=\pi$, we
obtain the local filtering enhancement of the separability criterion
introduced in ref.~{\cite{Zhang}}.

We stress that the new separability criteria induced by the
inequalities introduced in the present paper are, in principle,
practically implementable, since they involve easily computable
quantities related to the density matrix and its marginals. Future
work will be devoted to provide further examples and results along
the lines traced in the present contribution.

\ack The authors wish to thank G.~Marmo for invaluable human and
scientific support. C.~L.\ acknowledges the support of the project
CONQUEST, MRTN-CT-2003-505089.


\section*{References}

\begin{thebibliography}{99}

\bibitem{Einstein} Einstein A, Podolsky B, Rosen N 1935
{\it Phys. Rev.} {\bf 47} 777

\bibitem{Schr1} Schr\"odinger E 1935
{\it Naturwissenschaften} {\bf 23} 807, 823, 844

\bibitem{Schr2} Schr\"odinger E 1935
{\it Proc. Camb. Phil. Soc.} {\bf 31} 555; 1936
{\it ibidem} {\bf 32} 446

\bibitem{NiCh} Nielsen M A, Chuang I L 2000,
{\it Quantum Computation and Quantum Information} (Cambridge:
Cambridge University Press)

\bibitem{Bouwmeester} Bouwmeester D, Ekert A and Zeilinger A (Eds.)
2000, {\it The Physics of Quantum Information: Quantum Cryptography,
Quantum Teleportation and Quantum Computation} (New York: Springer)

\bibitem{Bruss} Bruss D 2002
{\it J. Math. Phys.} {\bf 43} 4237

\bibitem{Horos} Horodecki R, Horodecki P, Horodecki M, Horodecki K
2007 Quantum entanglement {\it Preprint} quant-ph/0702225

\bibitem{Werner} Werner R F 1989
{\it Phys. Rev. A} {\bf 40} 4277

\bibitem{Horo} Horodecki P 1997
{\it Phys. Lett. A} {\bf 233} 333

\bibitem{Horo96} Horodecki M, Horodecki P and Horodecki R 1996
{\it Phys. Lett. A} {\bf 223} 1

\bibitem{Peres96} Peres A 1996
{\it Phys. Rev. Lett.} {\bf 77} 1413

\bibitem{3H} Horodecki M, Horodecki P and Horodecki R 2006
{\it Open~Sys.~Inf.~Dyn.} {\bf 13} 103

\bibitem{Horo99} Horodecki M, Horodecki P 1999
{\it Phys. Rev. A} {\bf 59} 4206

\bibitem{Cerf} Cerf N J, Adami C, Gingrich R M 1999
{\it Phys. Rev. A} {\bf 60} 898

\bibitem{Nielsen} Nielsen M A, Kempe J 2001
{\it Phys. Rev. Lett.} {\bf 86} 5184

\bibitem{ReCr} Chen K, Wu L A 2003
{\it Quant.~Inf.~Comp.} {\bf 3} 193

\bibitem{CCN} Rudolph O 2002 Further results on the cross norm criterion for
separability {\it Preprint} quant-ph/0202121; Rudolph O 2003 {\it
Phys.~Rev.~A} {\bf 67} 032312

\bibitem{Zhang} Zhang C J, Zhang Y S, Zhang S,
Guo G C 2007 Entanglement detection beyond the cross-norm or
realignment criterion {\it Preprint} quant-ph/0709.3766

\bibitem{Schmidt} Peres A 1993 \emph{Quantum Theory: Concepts and
Methods} (Dordrecht: Kluwer Academic Publishers)

\bibitem{Horn} Horn R A, Johnson C R 1991 \emph{Topics in Matrix
Analysis}, (Cambridge: Cambridge University Press)

\bibitem{Schmidt_m} Terhal B M and Horodecki P 2000
{\it Phys.~Rev.~A} {\bf 61} 040301(R)
%
\bibitem{Eisert} Eisert J and Briegel J 2001
{\it Phys.~Rev.~A} {\bf 64} 022306

\bibitem{Lupo} Lupo C, Aniello P, Scardicchio A 2008 Bipartite quantum systems: on
the realignment criterion and beyond {\it Preprint}
quant-ph/0802.2019

\bibitem{Hyllus} G\"uhne O, Hyllus P, Gittsovich O and Eisert J 2007
{\it Phys. Rev. Lett.} {\bf 99} 130504


\end{thebibliography}
\end{document}